# Two-dimensional quantum transport of multivalley (111) surface state in topological crystalline insulator SnTe thin films

*Ryota Akiyama[§,\*], Kazuki Fujisawa, Tomonari Yamaguchi, and Shinji Kuroda[\*]*

Institute of Materials Science, University of Tsukuba, 1-1-1 Tennoudai, Tsukuba, 305-8573, Japan

Magneto-transport properties of (111)-oriented single-crystal thin films of SnTe were investigated. SnTe (111) thin films were epitaxially grown on a $BaF_2$ substrate by molecular beam epitaxy. By optimizing the growth conditions and the thickness of the films, the bulk carrier density could be reduced, making it possible to detect the surface transport. In the magneto-conductance (MC) measurement, a cusp-like feature around zero magnetic field was observed, which is attributed to the weak-antilocalization effect of the transport in the topological surface state. Detailed analysis of this negative MC reveals a reduced number of transport channels contributing to the surface transport, suggesting a strong coupling between Dirac valleys on the SnTe (111) surface, as a characteristic feature of the transport in the multivalley structure of TCI.

Topological insulators (TIs) have been attracting attention due to the emergence of the metallic state on their surface called topological surface state (TSS) exhibiting peculiar properties such as massless fermion on the Dirac-cone band dispersion and the spin-momentum locking.[1-3] Recent explorations for novel TI materials have opened the door to studies of a new type of TI, topological crystalline insulators (TCIs), in which the TSS is protected by the mirror reflection symmetry of the crystal structure,[4,5] instead of the time reversal symmetry in the conventional TIs.

SnTe was predicted to be a typical material of TCIs, with the mirror reflection symmetry with respect to the {100}, {111} and {110} planes of the rock-salt crystal structure. Experimentally, SnTe[6] and related mixed crystals (Pb,Sn)Se[7] have been confirmed to be TCIs with the observation of the Dirac-shape surface band by angle-resolved photoemission spectra (ARPES). In addition, the transport properties characteristic of the TSS, such as the Shubnikov-de Haas (SdH) oscillation, the Altshuler-Aronov-Spivak (AAS)/ Aharonov-Bohm (AB) effects, have been reported in nanowires,[8] SnTe (111) films grown on a conventional TI material $Bi_2Te_3$[9] and SnTe(100) films grown on a trivial insulator $BaF_2$.[10]

In the present study, we have investigated the electric transport in the TSS of SnTe (111) thin films grown on a $BaF_2$ substrate by molecular beam expitaxy (MBE).[11] Generally, transport experiments with SnTe have been regarded as a difficult task due to high-density holes in bulk originating from Sn vacancies, but we have succeeded in suppressing the contribution of the bulk transport by optimizing the growth conditions and the thickness of films. In the transverse magneto-conductance (MC) measurement, we observed a sharp cusp near zero field, which is attributed to the weak antilocalization (WAL) effect arising from the accumulation of $\pi$ Berry phase in the transport of carriers in the Dirac valleys on the SnTe (111) surface. An analysis of the WAL effect reveals a coherent coupling between these valleys, which is considered to be a characteristic feature of the surface state having the multivalley structure in a TCI.

Thin films of single-crystal SnTe were grown on a $BaF_2$ (111) substrate by MBE equipped with a compound source of SnTe. A $BaF_2$ substrate was introduced in a vacuum chamber after degreased using acetone and heated up to 300 °C for 1 hour for the thermal cleaning. Then SnTe was deposited on the surface at a substrate temperature $T_S$ = 250 °C. At this growth temperature, the single-crystal SnTe layer was epitaxially grown in the (111) orientation. The window of $T_S$ for the growth of the single-crystal layer was found to be relatively narrow; the deposition at a lower $T_S$ resulted in the formation of mixed regions grown both in (111) and (001) orientations. On the other hand, SnTe was not deposited at a higher $T_S$. At an optimal temperature of $T_S$ = 250 °C, we could obtain SnTe (111) epitaxial layer with a relatively low hole density, as will be shown later. The electric transport measurements were performed in the standard four-probe or van der Pauw configuration with electrodes fabricated by soldering indium on the surface of the SnTe layer. In the present article, we describe the result of the two films, samples A and B, with different thicknesses 46 nm and 74 nm, respectively, which were grown in exactly the same condition.

Figure 1a shows the temperature dependence of the resistivity of samples A and B. As shown in the figure, the both films exhibit metallic temperature dependence with an almost monotonic decrease in resistivity with lowering temperature, which was also reported in SnTe films grown on $Bi_2Te_3$.[9] The resistivity of sample A (46 nm thick) is larger than that of sample B (74 nm thick) by a factor of ~7 in the whole temperature range. In order to estimate the carrier density and mobility, we performed Hall measurement under a magnetic field up to 8 T. The Hall conductance $G_{xy}$ was deduced from the Hall resistance $R_{xy}$ and the sheet resistance $R_S$ by using Drude conductance tensor as shown below.

$$G_{xy} = \frac{R_{xy}}{R_{xy}^2 + R_S^2}, \quad (1)$$

As shown in Figure 1b, the deduced Hall conductance $G_{xy}$ exhibits a non-linear dependence on the applied magnetic field $B$. To fit these non-linear curves of $G_{xy}(B)$, we should use the Drude expression of the Hall conductivity,

$$G_{xy} = \frac{n_{2D} e \mu^2 B}{1 + (\mu B)^2}, \quad (2)$$

where $n_{2D}$ and $\mu$ represent the carrier density per unit area (2D carrier density) and the mobility, respectively. However, we could not achieve a satisfactory result of fitting using Eq. (2) assuming only one kind of carrier. In order to obtain a better fit, we need to use the following equation consisting of the sum of the conductivities of two kinds of carrier:[12]

$$G_{xy} = \frac{n_{s2D} e \mu_s^2 B}{1 + (\mu_s B)^2} + \frac{n_{b2D} e \mu_b^2 B}{1 + (\mu_b B)^2}, \quad (3)$$

where $n_{s2D}$ and $n_{b2D}$ represent the 2D carrier density and $\mu_s$ and $\mu_b$ represent the mobility of two kinds of carrier, respectively. As represented by green lines in Figure 1b, the results of fitting using eq 3 are quite well, and the values of the parameters obtained from the fitting are listed in

Table 1. The result of fitting shows that the both kinds are holes with different values of density and mobility. From this result, it is reasonably assumed that the one kind with a lower mobility and a higher density is the holes in bulk and the other kind with a higher mobility and a lower density is the holes on the surface.[13] The carrier density in sample A, in particular, that in bulk, 1.1 x 10$^{18}$ cm$^{-3}$ in the value converted into the density per unit volume (3D density), is the smallest value ever reported for SnTe films grown on a BaF$_2$ substrate.[10] From the obtained values of the 3D density in bulk, the position of the Fermi level is estimated roughly at about 22 meV or 185 meV below from the top of the valenceband in samples A, B, respectively, using the band model described in Ref. 13.[13] The significant reduction of the bulk hole density in sample A with a smaller thickness may possibly due to the depletion of holes at the region near the surface,[14] resulting in a relatively large contribution of the surface carriers to the electrical transport, in contrast to a still dominant contribution of the bulk carriers in sample B with a larger thickness.

Next, we discuss the quantum feature of the surface transport observed in the magnetoresistance (MR) measurement. We measured the angle-dependent MR under a magnetic field up to 8 T using a Quantum Design PPMS system equipped with a rotational sample holder. . From the obtained longitudinal (sheet) resistance $R_{xx}(B, \theta)$ at a magnetic field $B$ applied at an angle $\theta$ against the film surface (see inset in Figure 2a), we derived the magneto-conductance (MC) $\Delta G(B, \theta)$, a difference of the conductance at a given field $B$ from the value at zero field, defined as $\Delta G(B, \theta) = G(B, \theta) - G(0, \theta) = 1/R(B, \theta) – 1/R(0, \theta)$. In the MC curves shown from now on, we subtracted a component of the odd function of magnetic field, which originates from the voltage due to a small deviation of the alignment of the electrodes. In Figure 2, we plot $\Delta G(B, \theta)$ against $B$ at the respective values of $\theta$. As shown in the figure, the curves of $\Delta G(B, \theta)$ exhibit cusp-like behaviors, that is, a rapid decrease of the conductance with the application of a small magnetic field, for all the values of $\theta$. This MC is attributed to the WAL effect of the carrier transport.[15] The WAL effect is known as one of the features of the quantum transport in the diffusive transport regime, arising from the interference of wave function. In the transport of TSS, a shift of $\pi$ in the Berry phase arises from the spin-momentum locking effect for an electron (or a hole) traveling along a self-intersecting path, which causes a destructive interference of the wave function, leading to an enhancement of conductivity.[16] Upon the application of a magnetic field perpendicular to the closed path, this interference effect falls off, leading to a reduction of the conductivity.

The observed negative MC under a magnetic field perpendicular to the film plane ($\theta = 90°$) is consistent with the WAL effect in the transport of TSS. On the other hand, as shown in Figure 2, a negative MC was observed even under a magnetic field parallel to the film plane ($\theta = 0°$). This negative MC at $\theta = 0°$ could be attributed to the WAL effect of the three-dimensional (3D) transport of bulk carriers[17,18]. Even in a trivial system, the WAL effect arises from a destructive interference, similarly to the case of TSS, in the presence of a strong spin-orbit interaction. In the case of the 3D transport in bulk, the WAL effect should appear irrespective of the direction of a magnetic field.[19] Therefore, it can reasonably be assumed that the MC under an in-plane magnetic field reflects the WAL effect of only the 3D transport, while the MC under a perpendicular field reflects the WAL effect of both the 2D and 3D transports. If it is allowed to neglect the dependence of the 3D WAL effect on the applied magnetic field direction, the MC component arising from to the 2D WAL effect can be extracted by subtracting the MC under an in-plane magnetic field from that under a perpendicular field.[19] Figure 3 shows the result of the above subtraction, that is, $\Delta G(B, \theta) - \Delta G(B, 0°)$ is plotted against the component of magnetic field perpendicular to the film plane, $B \sin \theta$. The curves of the subtracted MC component decrease first with the increase of $B \sin \theta$, but it turns to an increase above a certain value of $B \sin \theta$, indicating the existence of both negative and positive components of MC. As seen in the figure, the decreasing parts of the MC curves of the respective values of $\theta$ collapse onto a universal curve, suggesting that the negative MC component originates from the 2D WAL effect of TSS.[19] On the other hand, the positive component may be attributed to the weak localization (WL) effect of the transport in the bulk subband formed at the surface, having a two-dimensional (2D) character, which we will discuss later.

According to a theory of the quantum correction of the 2D diffusive transport, the negative MC due to the WAL effect is given by the Hikami-Larkin-Nagaoka (HLN) equation[15]

$$\Delta\sigma_{2D} \equiv \sigma_{2D}(B) - \sigma_{2D}(0) = -\frac{\alpha e^2}{2\pi^2 \hbar}\left[\ln\left(\frac{\hbar}{4Bel_\phi^2}\right) - \psi\left(\frac{1}{2} + \frac{\hbar}{4Bel_\phi^2}\right)\right], \quad (4)$$

where $e$ is the charge of an electron, $\hbar$ is the Plank's constant, $l_\phi$ is the phase coherent length, $\psi(x)$ is the digamma function, and $\alpha$ represents a coefficient related to the number of coherent transport channels. Depending on the nature of the elastic scattering, the value of $\alpha$ per one transport channel should be 1, 0, or -0.5 for the orthogonal, unitary or symplectic class, respectively.[15] The 2D transport of TSS belongs to the symplectic class, giving $\alpha = -0.5$ per one transport channel. The subtracted MC $\Delta G(B, \theta) - \Delta G(B, 0°)$ can be fitted by eq 4 with $l_\phi$ and $\alpha$ as the fitting parameters. The experimental data can be fitted well, as represent the fitted curves by solid lines in Figure 3. The values of these parameters obtained from the fitting are $l_\phi = 230$ nm, $\alpha = -0.51$ for sample A and $l_\phi = 635$ nm, $\alpha = -0.34$, respectively. In the present case of the 2D transport of TSS, which belongs to the symplectic class, $\alpha = -0.5$ is expected theoretically for one transport channel.

In the TSS of SnTe (111), four valleys (Dirac cones) exist on the surface Brillouin zone (BZ), one in the $\overline{\Gamma}$ point and three in the $\overline{M}$ point, the four L points in the 3D BZ being projected along the [111] direction.[20,21] If these valleys on the both sides of the surface contribute to the transport as an independent channel, the total number of valleys counts as eight, $\alpha = -4$ thus being expected. On the other hand, the absolute values of $\alpha$ obtained from the fitting, $|\alpha| = 0.51$ for sample A and $|\alpha| = 0.34$ for sample B, are much smaller than the above expectation, which suggests that the number of effective transport channels is reduced due to a coupling between these valleys. If the carriers are scattered from one valley to another while keeping their coherence, these valleys are coupled and contribute to the transport as a single transport channel.[10,22] The coupling may be possible between the valleys on the same surface (intra-surface coupling), between the valleys on the other surface (inter-surface coupling), and between the valley on the surface and the state in bulk (surface-bulk coupling).[10] Following the above picture, the obtained value of $|\alpha| = 0.51$ for sample A can be interpreted as a result that all the valleys on the both surfaces couple into one single channel. On the other hand, a smaller value of $|\alpha| = 0.34$ for sample B, meaning the number of effective channels less than one, could not be explained only by the coupling between the valleys. It may suggest the contribution of a positive MC due to the WL effect in the bulk subband formed at the surface. The bulk band is split into 2D subbands due to a quantum confinement along the perpendicular direction of the layer and these bulk subbands may give rise to the WL effect, instead of the WAL effect, even in the case of topologically non-trivial material, as discussed in Ref. 23. Therefore, the observed MC is considered to be the superposition of the negative and positive MC component, arising from the WL and WAL effects in the surface and the bulk subband transport respectively. This would give an explanation for the inflected shape of the observed MC curve and an apparently smaller value of |α| than expected from the number of transport channels as a result of the cancellation between the positive and negative MC components.[23]

Then we investigated how the above features of the 2D transport vary with temperature of the measurement. In Figure 4, we plot the subtracted MC, $\Delta G(B, 90°) - \Delta G(B, 0°)$, at elevated temperatures from 5 to 30 K as a function of the component of magnetic field perpendicular to the film plane, $B \sin\theta$. As shown in the figure, the both positive and negative MC components are reduced rapidly by raising temperature. Similarly to the result at 4K, the negative MC component was fitted by HLN equation, with $l_\phi$ and $\alpha$ as the fitting parameters. The obtained values of $l_\phi$ and $\alpha$ are plotted against temperature in Figure 5. As shown in Figure 5a, $l_\phi$ decreases with increasing temperature $T$, exhibiting an almost linear dependence in the log-log plot, suggesting a power law relation as $l_\phi \propto T^{-\gamma}$. From the fitting to a linear dependence in Figure 5a, we obtained $\gamma = 0.51$ for samples A and $\gamma = 1.4$ for sample B. According to the theory of the diffusive transport[24,25], $l_\phi$ shows different power-law dependences on temperature by the mechanism of phase breaking scattering; $\gamma = 1$ should be given for the electron-phonon scattering, and $\gamma = 0.5$ should be given for the electron-electron scattering. The experimentally obtained value of $\gamma = 0.51$ in sample A suggests that the electron-electron scattering is the dominant mechanism of phase decoherence,[26]. On the other hand, the value $\gamma = 1.4$ in sample B does not match the value deduced from the theory. In the MC curves plotted in Figure 4, the negative MC component falls rapidly with raising temperature, making it difficult to distinguish from the positive MC component. Therefore, the fitting to the HLN equation may be affected by the superposition of the positive MC component. The temperature dependence of $\alpha$ shown in Figure 5b also exhibits different behaviors between the two samples. In the both samples, $|\alpha|$ decreases with the increase of temperature, but the decrease of $|\alpha|$ in sample A is almost monotonic, as shown in the figure. This decrease in $|\alpha|$ in sample A can be attributed to an enhanced coupling between different transport channels;[27] as the coherent scattering of carriers between transport channels rises by raising temperature, the coupling is enhanced, resulting in a decrease of $|\alpha|$. On the other hand, in sample B, $|\alpha|$ drops suddenly upon increasing temperature from 4 K to 4.5 K, but does not change much with the further increase of temperature. As discussed previously, even at the lowest temperature of 4 K, the value of $|\alpha|$ is already smaller than 0.5 as a result of the cancellation of the negative and positive MC components arising from WAL and WL.[23] With the increase of temperature, the coupling between the conduction channels is expected to be enhanced for both those contributing to the WAL and WL effects, which results in a further decrease in the apparent value of $|\alpha|$.

In conclusion, we have investigated the transport properties of the TSS in SnTe (111) thin films grown on a $BaF_2$ substrate by MBE. By controlling the growth condition, we have succeeded in fabricating epitaxial films grown only in the (111) orientation. In the MR measurement, we observed a negative MC near zero field, which is attributed to the WAL effect arising from $\pi$ shift in the Berry phase in the transport of TSS. By analyzing the dependence of the observed negative MR on the applied magnetic field direction, we extracted the 2D component of the MC and deduced the phase coherence length $l_\phi$ and the coefficient $\alpha$ related to the number of the transport channels from the fitting to the theoretical equation. The absolute value of $\alpha$ deduced from the fitting is much smaller than expected from the number of valleys in the TSS of SnTe (111), suggesting a strong coupling between the valleys, which is considered to be characteristic of the 2D transport of TSS with the multivalley electronic structure in TCI.

FIGURES

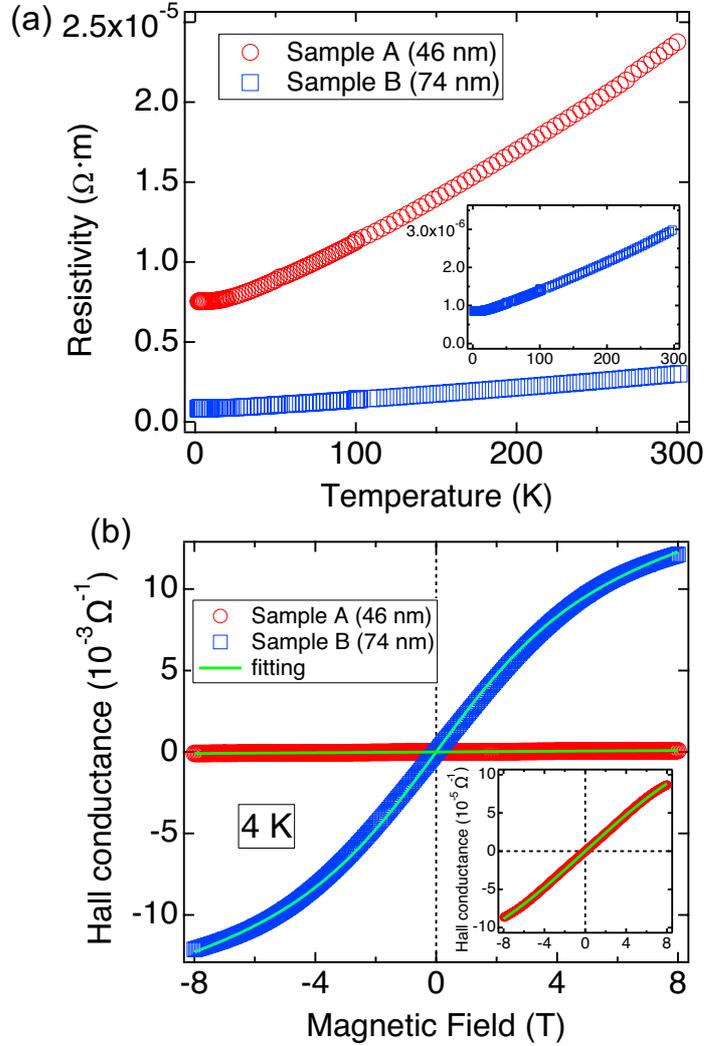

**Figure 1.** (a) The temperature dependence of resistivity of samples A, B. (b) The Hall conductance $G_{xy}$ as a function of the applied magnetic field in samples A, B. The measurement was performed at 4 K. Green lines represent the result of fitting using eq 3, assuming the two kinds of carrier in bulk and on the surface.

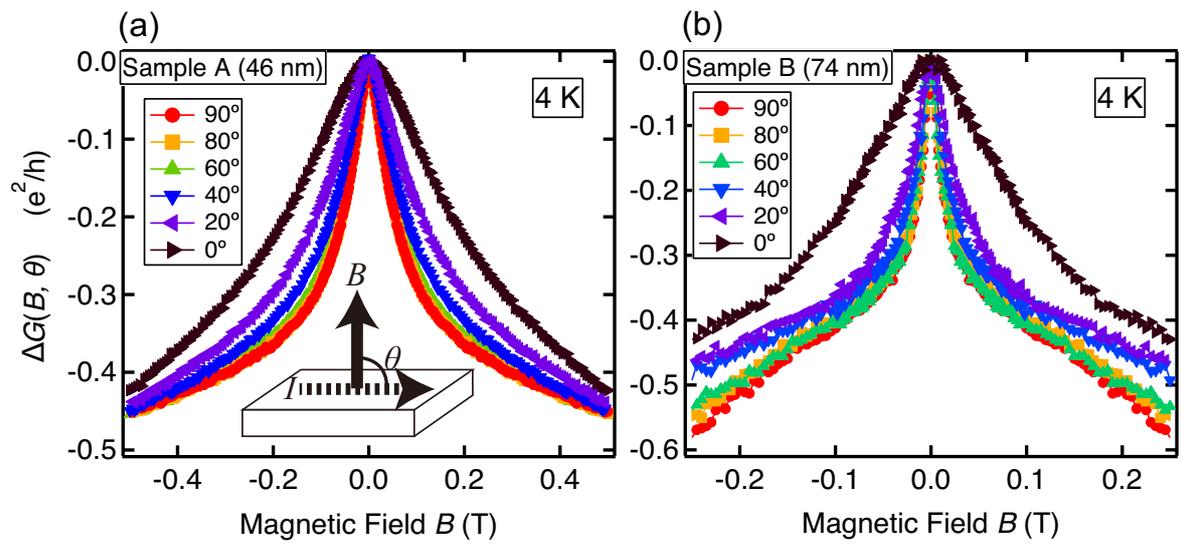

**Figure 2.** Magneto-conductance (MC) $\Delta G(B, \theta) = G(B, \theta) - G(0, \theta)$ under the application of magnetic field $B$ at the respective angles $\theta$. (a) and (b) represent the data of samples A, B taken at 4 K.

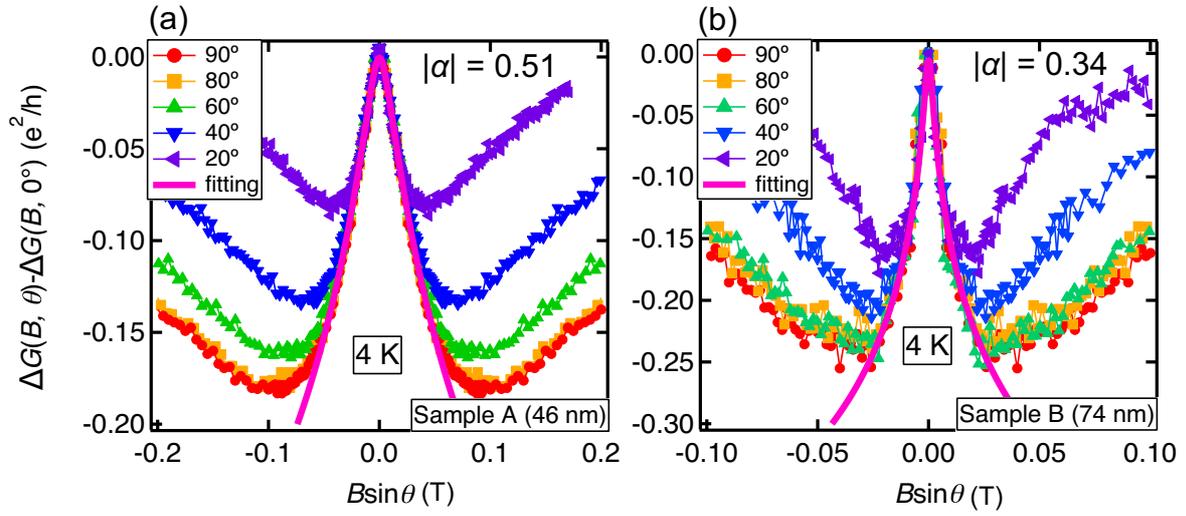

**Figure 3.** The 2D MC component calculated by the subtraction $\Delta G(B, \theta) - \Delta G(B, 0°)$ is plotted as a function of the magnetic field component perpendicular to the film plane, $B \sin \theta$. (a) and (b) represent the data of samples A, B taken at 4 K. Red lines represent the result of fitting of the negative MC component to the HLN equation (eq 4).

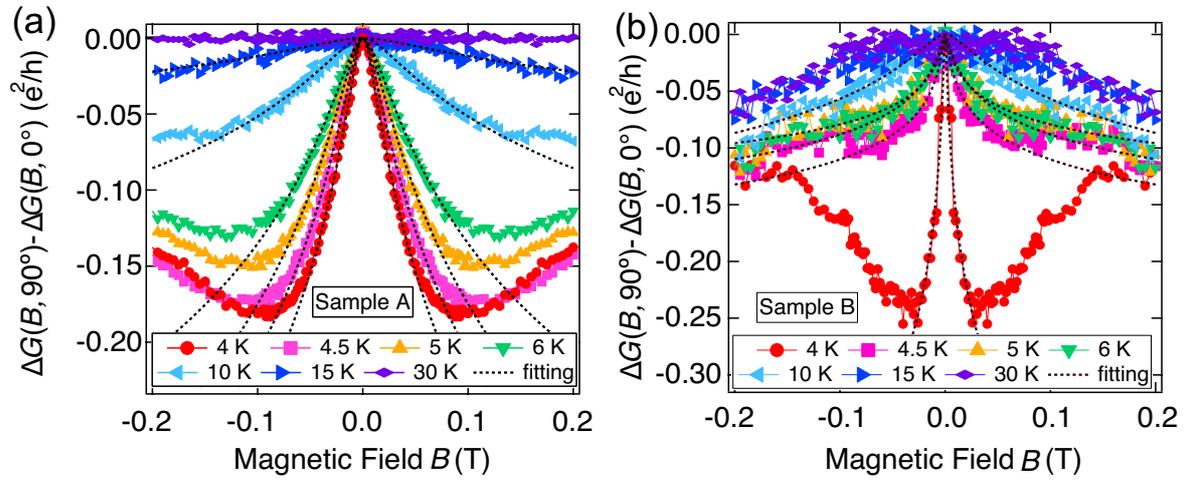

**Figure 4.** The 2D MC component calculated by the subtraction $\Delta G(B, 90°) - \Delta G(B, 0°)$ as a function of the applied magnetic field at the respective temperatures 4 - 30 K. (a) and (b) represent the data of samples A, B. Dotted lines represent the result of fitting of the negative MC component.

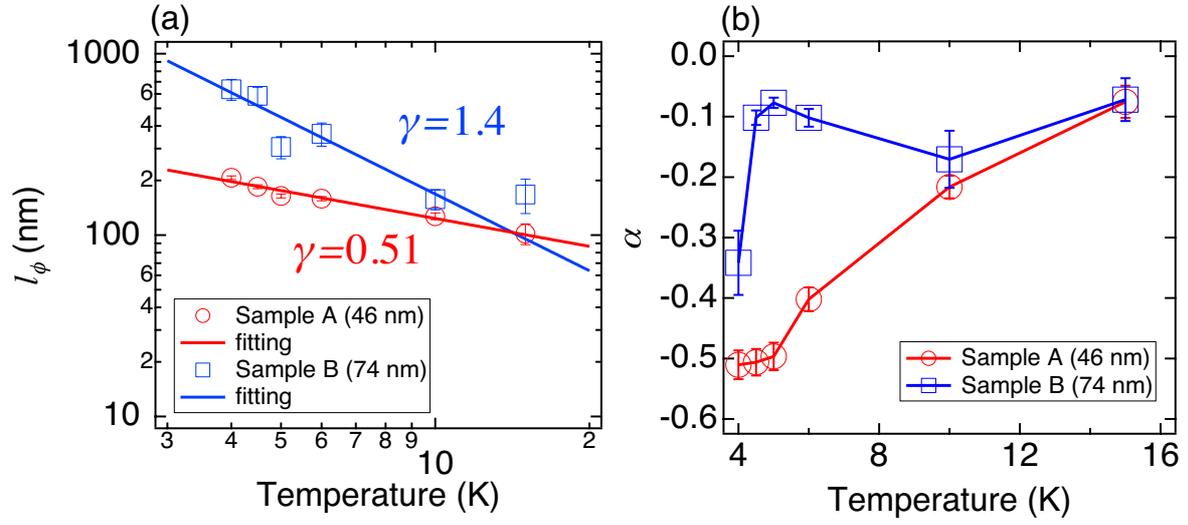

**Figure 5.** The temperature dependence of (a) the phase coherence length $l_\phi$ and (b) the coefficient $\alpha$ related to the number of coherent transport channels, obtained from the fitting of the MC data in Figure 4 to the HLN equation. The temperature dependence of $l_\phi$ was fitted to a power law relation of $l_\phi \propto T^{-\gamma}$ with the power $\gamma$ as a fitting parameter.

**Table 1.** The carrier density per unit area (2D density) and mobility of the two kinds of carriers obtained from the fitting of the curves of Hall conductance $G_{xy}$ as a function of magnetic field $B$ (Figure 1b). It is assumed that the one kind of carrier (Carrier I) with a lower mobility and a higher density is the carriers in bulk and the other kind of carrier (Carrier II) with a higher mobility and a lower density is the carriers on the surface. For Carrier I, which is assumed to be the carriers in bulk, the values of density converted in the density per unit volume (3D density) are also shown.

|  | Carrier I (bulk) | | | Carrier II (surface) | |
| --- | --- | --- | --- | --- | --- |
|  | 2D density ($cm^{-2}$) | 3D density ($cm^{-3}$) | Mobility ($cm^2$/V sec) | 2D density ($cm^{-2}$) | Mobility ($cm^2$/V sec) |
| Sample A (46nm) | $5.1 \times 10^{12}$ | $1.1 \times 10^{18}$ | 334 | $6.0 \times 10^{11}$ | 561 |
| Sample B (74nm) | $4.8 \times 10^{14}$ | $6.5 \times 10^{19}$ | 369 | $3.7 \times 10^{11}$ | 1565 |

## ASSOCIATED CONTENT

**Supporting Information**.

Figure S1: The fitting results by the empirical equation about the temperature dependence of resistivity. Figure S2: The temperature dependence of the Hall coefficient.

## AUTHOR INFORMATION

**Corresponding Author**

*E-mail: (R. A) akiyama@surface.phys.s.u-tokyo.ac.jp; (S. K) kuroda@ims.tsukuba.ac.jp

**Present Addresses**

§ Department of Physics, Faculty of Science, The University of Tokyo.

## ACKNOWLEDGMENT

The authors would like to thank T. Koyano in the Cryogenics Division of Research Facility Center for Science and Technology, University of Tsukuba for PPMS measurements. This work is partially supported by Grant-in-Aids for Scientific Research (KAKENHI) from the Japan Society for the Promotion of Science, CASIO Science Promotion Foundation, and Sumitomo Electric Industries Group CSR Foundation.

**Supporting Information**

**Two-dimensional quantum transport of multivalley (111) surface state in topological crystalline insulator SnTe thin films**

*Ryota Akiyama*[§,*], *Kazuki Fujisawa, Tomonari Yamaguchi, and Shinji Kuroda*[*]

Institute of Materials Science, University of Tsukuba, 1-1-1 Tennoudai, Tsukuba, 305-8573, Japan

*E-mail: (R. A) akiyama@surface.phys.s.u-tokyo.ac.jp; (S. K) kuroda@ims.tsukuba.ac.jp

§ Present address: Department of Physics, Faculty of Science, The University of Tokyo.

In order to elucidate a scattering mechanism of carriers, we analyzed the temperature dependence of the resistivity by fitting using the following empirical equation[1,2];

$$R = R_0 + \alpha \exp\left(-\frac{\theta}{T}\right) + \beta T^2 \qquad S1$$

where the second term represents the resistivity caused by the phonon scattering and the third term does that caused by the electron-electron (e-e) scattering. As shown in Figure S1a, b, the experimental data could be fitted well by eq S1. The values of the parameters obtained from the fitting are listed in Table S1. As the contributions of the respective terms of the phonon scattering and the e-e scattering shown in Figure S1c, d, in the both samples, the contribution of the e-e scattering is larger than that of the phonon scattering at low temperatures less than about 20 K, but the dominance of the e-e scattering is more significant in sample A. This seems to be consistent with the conclusion that the e-e scattering is the dominant mechanism of phase coherence in sample A, which was derived from the temperature dependence of the coherence length $l_\phi$ shown in Figure 5a. Figure S2 shows the temperature dependence of the Hall coefficient, which was deduced from the Hall resistance at 1 T. As shown in the figure, the Hall coefficient varies with temperature only gradually, without a change in sign in the both samples.

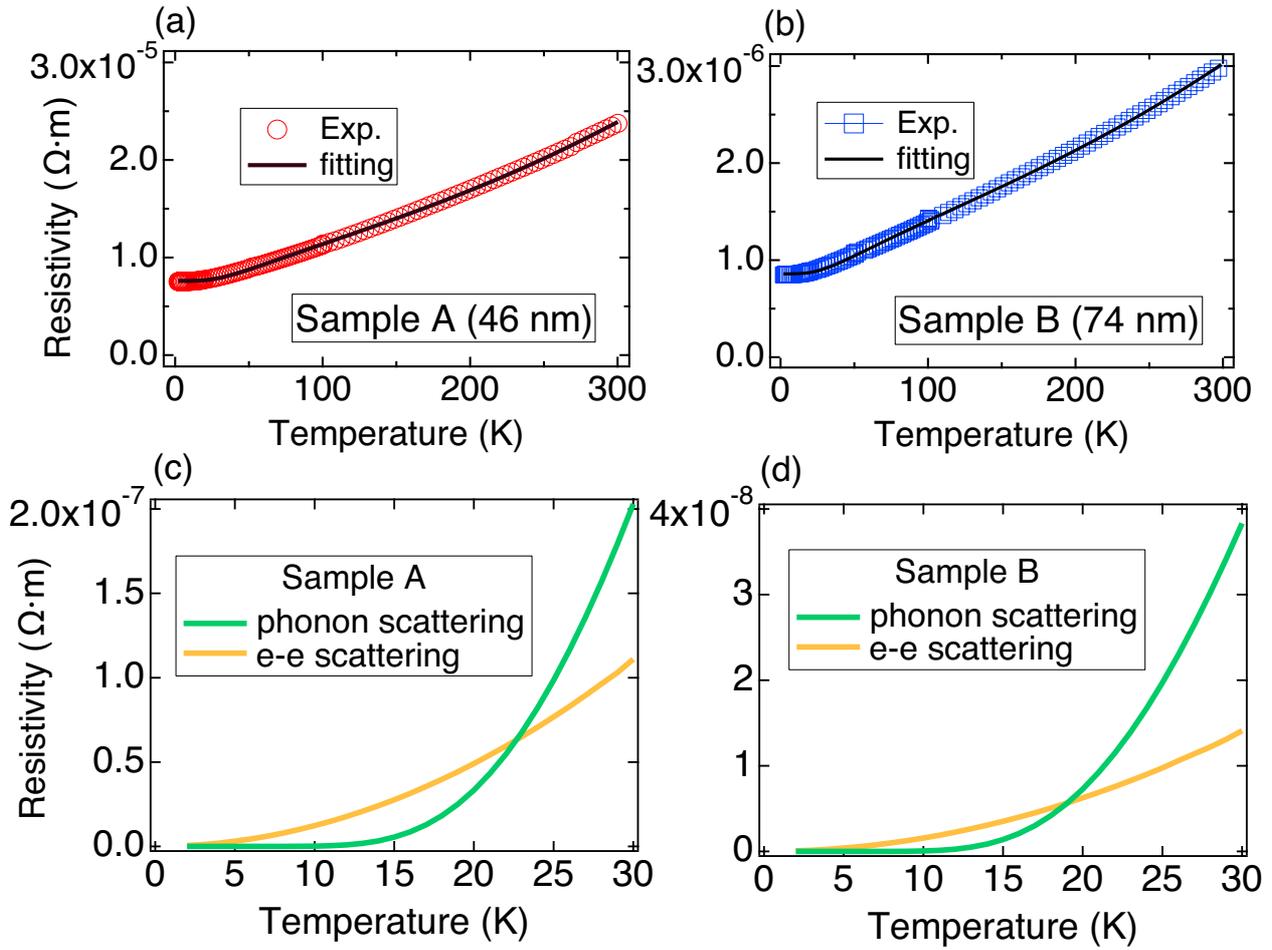

Figure S1: (a), (b) The temperature dependence of resistivity and the result of fitting using the empirical equation for samples A, B. Blue and red marks represent the experimental data and black lines represent the fitted curve. (c), (d) The contributions of the second term (phonon scattering) and of the third term (e-e scattering) in the fitted curve are shown by green and orange lines, respectively.

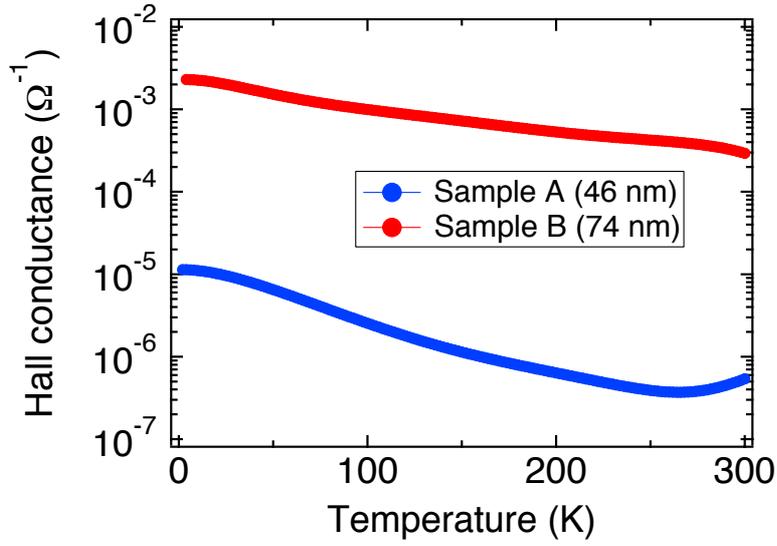

Figure S2: The temperature dependence of the Hall coefficient in samples A, B.

Table S1: The parameters obtained from the fitting to eq S1.

|  | $R_0$ (Ω·m) | $\alpha$ (Ω·m) | $\theta$ (K) | $\beta$ (Ω·m·K$^{-2}$) |
| --- | --- | --- | --- | --- |
| Sample A | $7.60 \times 10^{-6}$ | $7.53 \times 10^{-6}$ | 108.4 | $1.23 \times 10^{-10}$ |
| Sample B | $8.59 \times 10^{-7}$ | $1.10 \times 10^{-6}$ | 99.7 | $1.56 \times 10^{-11}$ |